\documentclass[aps, showpacs, showkeys,nofootinbib,floatfix]{revtex4}

\usepackage{amssymb}
\usepackage{amsmath}
\usepackage{graphicx}
\usepackage{hyperref}

\begin{document}

\title{The Real Solution to Scalar Field Equation in 5D Black String Space}

\author{Molin Liu}
\email{mlliudl@student.dlut.edu.cn}
\author{Hongya Liu}
\email{hyliu@dlut.edu.cn}
\author{Feng luo}
\author{Lixin Xu}

\affiliation{School of Physics and Optoelectronic Technology,
Dalian University of Technology, Dalian, 116024, P. R. China}

\begin{abstract}
After the nontrivial quantum parameters $\Omega_{n}$ and quantum
potentials $V_{n}$ obtained in our previous research, the
circumstance of a real scalar wave in the bulk is studied with the
similar method of Brevik (2001). The equation of a massless scalar
field is solved numerically under the boundary conditions near the
inner horizon $r_{e}$ and the outer horizon $r_{c}$. Unlike the
usual wave function $\Psi_{\omega l}$ in 4D, quantum number $n$
introduces a new functions $\Psi_{\omega l n}$, whose potentials
are higher and wider with bigger n. Using the tangent
approximation, a full boundary value problem about the
Schr$\ddot{o}$dinger-like equation is solved. With a convenient
replacement of the 5D continuous potential by square barrier, the
reflection and transmission coefficients are obtained. If extra
dimension does exist and is visible at the neighborhood of black
holes, the unique wave function $\Psi_{\omega l n}$ may say
something to it.
\end{abstract}



\maketitle

\section{Introduction}
In 1974 \cite{Hawking1} and 1975 \cite{Hawking2} Stephen Hawking
published his analysis of the effects of gravitational collapse on
quantum fields, and predicted that black holes are not perfect
black, but radiate thermally and eventually explode, named Hawking
radiation. Since then, many people have used various methods and
techniques to research black hole through the particles radiating
from it, such as the simple Klein-Gordon particles and Dirac
particles (for some early works, see Damour and Ruffini
\cite{Damour} and Chandrasekhar \cite{Chandrasekhar}
respectively). Recently, as for the former scalar field, Higuchi
et al. \cite{Higuchi} and Grispino et al. \cite{Grispino} gave the
scalar field solution outside a Schwarzschild black hole, Brady et
al. \cite{Brady}, Brevik et al. \cite{Brevik} and Tian et al.
\cite{Tian} studied the Schwarzschild-de Sitter case and Guo et
al. \cite{Guo} made further studies in the
Reissner-Nordstr$\ddot{o}$m-de Sitter case and so on.

Regarding higher dimensional background, in order to avoid the
interactions beyond any acceptable phenomenological limits, people
assume that standard model fields (such as fermions, gauge bosons,
Higgs fields) are confined on a ($3+1$) dimensional hypersurface
(3-brane) without accessing along the transverse dimensions. The
branes are embedded in the higher dimensional space (bulk), in
which only gravitons and scalar particles without charges could
propagate under standard model gauge group. In this paper, the
massless scalar field is chosen. Meanwhile, there are also many
works (for a review with large extra dimensions see Ref.
\cite{Kantis}) focusing on Hawking radiation such as Kanti, Frolov
and Harris et al. \cite{Kanti2} \cite{Frolov} \cite{Kanti4}
\cite{Harris} \cite{Kanti3}, in which they employed both
analytical and numerical techniques to calculate various
particles' greybody factors and different energy emission rates on
the brane and in the bulk outside a $(4+n)$-dimensional
Schwarzschild black holes and $(4+n)$-dimensional Schwarzschild-de
Sitter black holes. These are two kinds of small higher
dimensional black holes with horizon $r_{H}\ll L$, where $L$ is
the size of the extra dimensions. Such small black holes can be
treated safely as classical objects.

That the world may have more than four dimensions is due to Kaluza
(1921) \cite{Kaluza} and Klein (1926) \cite{Klein}, who realized
that a 5D manifold could be used to unify general relativity with
Maxwell's theory of electromagnetism. After that, many people are
more interested in the higher dimensional gravitational theory.
Here, we consider the Space-Time-Matter (STM) theory presented by
Wesson and co-workers \cite{Wesson} \cite{Overduin}. This theory
is distinguished from the classical Kaluza-Klein theory for a
non-compact fifth dimension, the 4D source is reduced from an
empty 5D manifold. Because of this, the STM theory is also called
induced matter theory and the effective 4D matter is called
induced matter. That is, in STM theory, the 5D manifold is
Ricci-flat while the 4D hypersurface is curved by the 4D induced
matter. Mathematically, this approach is supported by Campbell's
theorem which states that any analytical solution of N-dimensional
Einstein equations with a source can be locally embedded in an (N
+ 1)-dimensional Ricci-flat manifold \cite{Campbell}. In the
framework of STM, people studied many astrophysical implications
such as Perihelion Problem \cite{Lim}, Kaluza-Klein Solitons
\cite{Billyard} \cite{Liu_1}, Black hole \cite{Liu222}, Solar
System Tests \cite{Liu333} and so on. So far as we know, except
for our previous work \cite{Liu00}, no one has researched the
radiation of 5D black holes in STM theory before.

This paper is organized as follows: In Section II, the 5D black
string metric and the time-dependent radial equation about
$R_{\omega}(r,t)$ are given. In section III, by a tortoise
coordinate transformation, the radial equation becomes a
Schr$\ddot{o}$dinger-like equation. According to the boundary
condition and the tangent approximation, a full numerical solution
is presented. In section VI, with replacing the real potential
barriers around black hole by square barriers, the reflection and
transmission coefficients are naturally obtained. Section V is a
conclusion.

We adopt the signature $(+, -, -, -, -)$ and put $\hbar$, $c$ ,and
$G$ equal to unity. Lowercase Greek indices $\mu,\nu\ldots$ will
be taken to run over $0, 1, 2, 3$ as usual, while capital indices
A, B, C, $\ldots$ run over all five coordinates $(0, 1, 2, 3, 4)$.
\section{The Klein-Gordon Equation in the 5D Schwarzschild-de Sitter solution}
Within the framework of STM theory, a class of exact 5D solutions,
presented by Mashhoon, Wesson and Liu \cite{Liu1}
 \cite{Mashhoon} \cite{Wesson},
describes a 5D black hole.  The line element takes the form
\begin{equation}
dS^{2}=\frac{\Lambda
\xi^2}{3}\left[f(r)dt^{2}-\frac{1}{f(r)}dr^{2}-r^{2}\left(d\theta^2+\sin^2\theta d\phi^2\right)\right]-d\xi^{2}. \label{eq:5dmetric}%
\end{equation}
In this metric
\begin{equation}
f(r)=1-\frac{2M}{r}-\frac{\Lambda}{3}r^2,\label{f-function}
\end{equation}
where $\xi$ is the open non-compact extra dimension coordinate,
$\Lambda$ is the induced cosmological constant and $M$ is the
central mass. The part of this metric inside the square bracket is
exactly the same line-element as the 4D Schwarzschild-de Sitter
solution, which is bounded by two horizons
--- an inner horizon (black hole horizon) and an outer horizon (one may call this cosmological horizon). This metric
(\ref{eq:5dmetric}) satisfies the 5D vacuum equation $R_{AB}=0$
and, therefore, there is no cosmological constant when viewed from
5D. However, when viewed from 4D, there is an effective
cosmological constant $\Lambda$. So one can actually treat this
$\Lambda$ as a parameter which comes from the fifth dimension.
This solution has been studied in many works \cite{Mashhoon11}
\cite{Wesson_1} \cite{Liu_2} \cite{Mashhoon_1} focusing mainly on
the induced constant $\Lambda$, the extra force and so on.

We redefine the fifth dimension in this model,
\begin{equation}
\xi=\sqrt{\frac{3}{\Lambda}}e^{\sqrt{\frac{\Lambda}{3}}y}.\label{replacement}
\end{equation}
Then we use (\ref{eq:5dmetric}) $\sim$ (\ref{replacement}) to
build up a RS type brane model in which one brane is at $y=0$, and
the other brane is at $y=y_{1}$. Hence the fifth dimension becomes
finite. It could be very small as RS I brane model \cite{Randall2}
or very large as RS II model \cite{Randall1}. It has been noted
that the relevancy between STM and brane-worlds theories and some
embedding of 5D solutions to brane models are studied in
\cite{Ponce} \cite{Seahra} \cite{Liu_plb} \cite{Ping}. For the
present brane model, when viewed from a ($\xi$ or $y$ =
$constant$) hypersurface, the 4D line-element represents exactly
the Schwarzschild-de Sitter black hole. However, when viewed from
5D, the horizon does not form a 4D sphere --- it looks like a
black string lying along the fifth dimension. Usually, people call
the solution to the 5D equation $^{(5)}G_{AB}$ = $\Lambda _{5}$$
^{(5)}g_{AB}$ ($\Lambda_{5}$ is the 5D cosmological constant) as
the 5D Schwarzschild-de Sitter solution. Therefore, to distinguish
it, we call the solution (\ref{eq:5dmetric}) a black string, or
more precisely, a 5D Ricci-flat Schwarzschild-de Sitter solution.

In our previous paper \cite{Liu00}, we have introduced a massless
scalar field to stabilizing this black string brane model.
Considering a single mode of the scalar field, the wave function
for this mode may reach the maximum value but keep smooth and
finite at the brane. Hence, a steady standing wave is constructed.
A suitable superposition of some of the quantized and continuous
components of $L(y)$, which is one component of scalar field, may
provide a wave function which is very large at $y=0$ and drop
rapidly for $y \neq 0$. Naturally, a practical 3-brane is formed
at the $y=0$ hypersurface.

With the redefinition (\ref{replacement}), the metric
(\ref{eq:5dmetric}) can be rewritten as
\begin{equation}
dS^{2}=e^{2\sqrt{\frac{\Lambda}{3}}y}\left[f(r)dt^{2}-\frac{1}{f(r)}dr^{2}-r^{2}\left(d\theta^2+\sin^2\theta d\phi^2\right)-dy^{2}\right],\label{eq:5dmetric-y}%
\end{equation}
where $y$ is the new fifth dimension. Expression
(\ref{f-function}) can be recomposed as follows
\begin{equation}
f(r)=\frac{\Lambda}{3r}(r-r_{e})(r_{c}-r)(r-r_{o}). \label{re-f function}%
\end{equation}

The singularity of the metric (\ref{eq:5dmetric-y}) is determined
by $f(r)=0$. Here we only consider the real solutions. The
solutions to this equation are inner horizon $r_{e}$, outer
horizon $r_{c}$ and a negative solution $r_{o}=-(r_{e}+r_{c})$.
The last one has no physical significance, and $r_{c}$ and $r_{e}$
are given as

\begin{equation}
\left\{
\begin{array}{c}
r_{c} = \frac{2}{\sqrt{\Lambda}}\cos\eta ,\\
r_{e} = \frac{2}{\sqrt{\Lambda}}\cos(120^\circ-\eta),\\
\end{array}
\right.\label{re-rc}
\end{equation}
where $\eta=\frac{1}{3}\arccos(-3M\sqrt{\Lambda})$ with $30^\circ
\leq\eta\leq 60^\circ$. The real physical solutions are accepted
only if
 $\Lambda$ satisfy $\Lambda M^2\leq\frac{1}{9}$ \cite{Liu1}.

Following the work \cite{Lim}, the massless scalar field $\Phi$ in
the 5D black string space, satisfies the Klein-Gordon equation
\begin{equation}
\square\Phi=0,\label{Klein-Gorden equation}%
\end{equation}
where $ \square=\frac{1}{\sqrt{g}}\frac{\partial}{\partial
x^{A}}\left(\sqrt{g}g^{AB}\frac{\partial}{\partial{x^{B}}}\right)\label{Dlb}
$  is the 5D d'Alembertian operator. We suppose that the separable
solutions to Eq. (\ref{Klein-Gorden equation}) are of the form
\begin{equation}
\Phi=\frac{1}{\sqrt{4\pi\omega}}\frac{1}{r}R_{\omega}(r,t)L(y)Y_{lm}(\theta,\phi),\label{wave
function}
\end{equation}
where $R_{\omega}(r,t)$ is the radial time-dependent function,
$Y_{lm}(\theta,\phi)$ is the usual spherical harmonic function,
and $L(y)$ is the function about the fifth dimension. The
differential equation of $R_{\omega}(t, r)$ is
\begin{equation}
-\frac{1}{f(r)} r^2\frac{\partial^2}{\partial
 t^2}\left(\frac{R_{\omega}}{r}\right)+\frac{\partial}{\partial r}\left(r^2
 f(r)\frac{\partial}{\partial{r}}\left(\frac{R_{\omega}}{r}\right)\right)-\left[\Omega r^2+l(l+1)\right]\frac{R_{\omega}}{r}=0,\label{radius-t-equation}
\end{equation}
where $\Omega$ is a constant which is adopted to separate
variables $(t, r, \theta, \phi)$. Meanwhile, in our previous study
\cite{Liu00}, we have solved successfully the evolution equation
$L(y)$ in branes model. Its norm $|L(y)|^2$, which represents the
probability of finding the massless scalar particle, gets an
extremum on the branes. According to the standing wave condition
in the bulk, the spectrum of $\Omega$ is broken into two parts:
one is the continuous spectra below $\frac{3}{4}\Lambda$ and the
other is the discrete spectra above $\frac{3}{4}\Lambda$. The
quantum parameter $\Omega_{n}$ is
\begin{equation}\label{quan-Omega}
    \Omega_{n}=\frac{n^2\pi^2}{y_{1}^2}+\frac{3}{4}\Lambda,
\end{equation}
where $n=1, 2, 3 \cdots$ and $y_{1}$ is the thickness of the bulk.

\section{The Boundary Value Problem in the bulk}
\subsection{The Schr$\ddot{o}$dinger-like Equation}
A more important fact is the aspect of radial direction. In Eq.
(\ref{radius-t-equation}) time variable can be eliminated by the
Fourier component $e^{-i \omega t}$ via
 \begin{equation}
R_{\omega}(r,t)\rightarrow\Psi_{\omega l n}(r) e^{-i\omega t},
 \end{equation}
 where the subscript $n$ presents a new wave function unlike the usual 4D case $\Psi_{\omega l}$ \cite{Brevik}.  Eq. (\ref{radius-t-equation}) can be rewritten as%
 \begin{equation}
 \left[-f(r)\frac{d}{dr}(f(r)\frac{d}{dr})+V(r)\right]\Psi_{\omega
 l n}(r)=\omega^2\Psi_{\omega l n}(r),\label{radius equ. about r}
 \end{equation}
whose potential function is given by%
\begin{equation}
V(r)=f(r)\left[\frac{1}{r}\frac{df(r)}{dr}+\frac{l(l+1)}{r^2}+\Omega\right].\label{potential-of-r}
\end{equation}

Now we introduce the tortoise coordinate%
\begin{equation}
x=\frac{1}{2M}\int\frac{dr}{f(r)}.\label{tortoise }
\end{equation}
The tortoise coordinate can be expressed with the surface gravity
as follows
\begin{equation}
x=\frac{1}{2M}\left[\frac{1}{2K_{e}}\ln\left(\frac{r}{r_{e}}-1\right)-\frac{1}{2K_{c}}\ln\left(1-\frac{r}{r_{c}}\right)+\frac{1}{2k_{o}}\ln\left(1-\frac{r}{r_{o}}\right)\right],\label{tor-grav-sf}
\end{equation}
where%
\begin{equation}
K_{i}=\frac{1}{2}\left|\frac{df}{dr}\right|_{r=r_i}.
\end{equation}
Explicitly, we have%
\begin{eqnarray}
  K_{e}=\frac{(r_{c}-r_{e})(r_{e}-r_{o})}{6r_{e}}\Lambda, \\
  K_{c}= \frac{(r_{c}-r_{e})(r_{c}-r_{o})}{6r_{c}}\Lambda,\\
  K_{o}= \frac{(r_{o}-r_{e})(r_{c}-r_{o})}{6r_{o}}\Lambda.
\end{eqnarray}
So under the tortoise coordinate transformation (\ref{tortoise }),
the
radial equation (\ref{radius equ. about r}) can be written as%
\begin{equation}
\left[-\frac{d^2}{dx^2}+4M^2V(r)\right]\Psi_{\omega l
n}(x)=4M^2\omega^2\Psi_{\omega l n}(x),\label{radius-equation}
\end{equation}
which is of the form of Schr$\ddot{o}$dinger equation in quantum
mechanics. Because there are two different coordinates --- $r$ and
$x$ in it, people usually call it Schr$\ddot{o}$dinger-like
equation. The incoming or outgoing particle flow between inner
horizon $r_{e}$ and outer horizon $r_{c}$ is reflected and
transmitted by the potential $V(r)$. Substituting the quantum
parameters $\Omega_{n}$ (\ref{quan-Omega}) into Eq.
(\ref{potential-of-r}), the quantum potentials are obtained as
follows
\begin{equation}\label{quan-potential}
    V_{n}(r)=f(r)\left[\frac{1}{r}\frac{df(r)}{dr}+\frac{l(l+1)}{r^2}+\frac{n^2\pi^2}{y_{1}^2}+\frac{3}{4}\Lambda\right].
\end{equation}
This means that the potential $V(r)$ (\ref{quan-potential})
determines the evolution of the field $\Psi_{\omega l n}$.
\subsection{The numerical solution}
Near the horizons $r_{e}$ and $r_{c}$, the tortoise coordinate
$x\longrightarrow\pm\infty$. According to Eq. (\ref{re-f
function}) and Eq. (\ref{quan-potential}), the boundary conditions
are as follows
\begin{equation}\label{Boundary-potential}
    V(r_{e})=V(r_{c})=0.
\end{equation}
So near the horizons, Eq. (\ref{radius-equation}) reduces to
\begin{equation}\label{Boundary-Equ}
    \left[\frac{d^2}{dx^2}+4M^2\omega^2\right]\Psi_{\omega l
    n}(x)=0,
\end{equation}
where the potential vanishes. Obviously, its solutions are $e^{\pm
i2M\omega x}$. In this paper, taking into account only real field,
we choose the solution \cite{Brevik}
\begin{equation}\label{Boundary-conditions}
   \Psi_{\omega l n}=\cos(2M\omega x)
\end{equation}
as a boundary condition near the two horizons (\ref{re-rc}).

\begin{figure}
\centering
\includegraphics[width=3.5in]{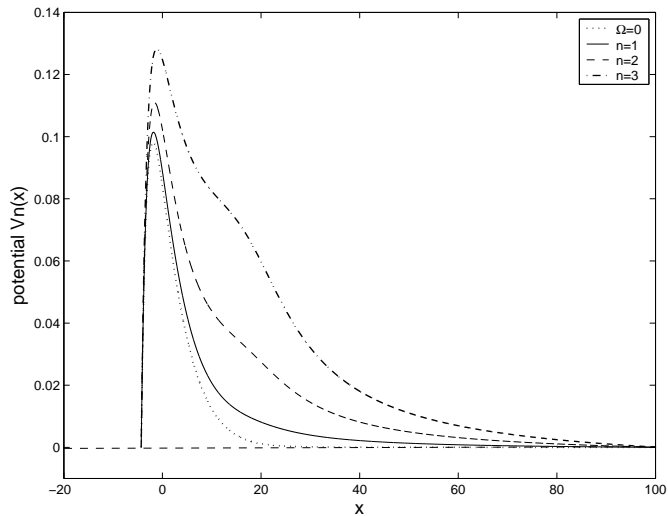}
\caption{The first three potential barriers $V_{n}(x)$ of 5D black
string with $n=1$ (solid), $n=2$ (dashed) and $n=3$ (dash-dot). We
use $l=1$, $M=1$, $\Lambda=10^{-3}$ and $y_{1}=10^{3/2}$ (a very
large 5th dimension). Meanwhile, the potential barrier of the
corresponding 4D solution ($\Omega=0$) is plotted for
comparison.}\label{fig:potential}
\end{figure}

Because there are two coordinates --- radial coordinate $r$ and
tortoise coordinate $x$ coupled in Eq.(\ref{radius-equation}), the
source transformation expression (\ref{tor-grav-sf}) is too
complex to be solved. In order to solve Eq.
(\ref{radius-equation}) conveniently, it is necessary to use an
approximate method for transition
\begin{equation}\label{1}
    \tilde{r}(x)=f(x).
\end{equation}
As far as we know, there are two methods used frequently  --- one
is tangent approximation \cite{Brevik}, and the other is
polynomial approximation \cite{Tian}. The former method is
convenient for theoretical analysis and is adopted here. The
latter one involves too many polynomials approximation and we do
not consider it in this paper. To the parameter $\Lambda$, we can
always find an appropriate approximate method from the above. For
a widely separated horizons model, we adopt the same value
$\Lambda=10^{-3}$ as appeared in \cite{Brevik} \cite{Tian}. Then
putting $\Lambda$ into the Eq. (\ref{re-rc}), we find the inner
horizon $r_{e}=2.00268M$ and the outer horizon $r_{c}=53.7435M$.
We employ the useful tangent approximation
\begin{equation}
  \tilde{r}(x)=\frac{1}{b}\arctan\left[\frac{1}{10}(x-18)\right]+d,\label{tangent-appr}
\end{equation}
in which $b=2.6/(r_{c}-r_{e})$ and $d=(r_{c}+r_{e})/2-3$ are two
parameters to linearly fit $r$ and $x$. Thus, the potentials
$V_{n}(r)$ (\ref{quan-potential}) can be converted into
$V_{n}(x)$, which are plotted in Fig. \ref{fig:potential}. Clearly
those potentials are highly localized near $x=0$ and fall off
exponentially at both inner horizon $r_{e}(x)$ and outer horizon
$r_{c}(x)$. It is characteristic in our higher dimensional
scenario that potentials are higher with bigger quantum number
$''n''$.

\begin{center}
\begin{figure}
\centering
\includegraphics[width=3.5in]{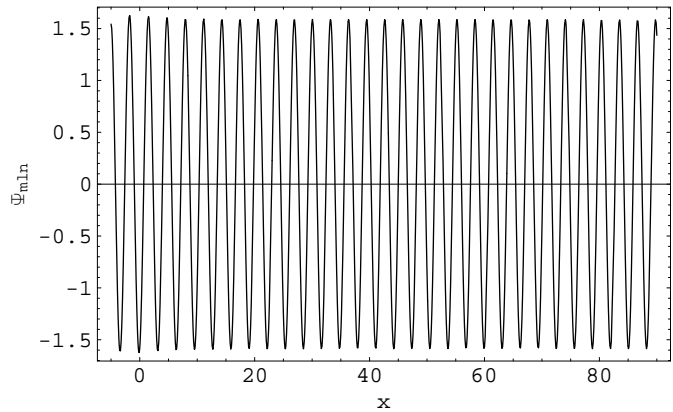}
\caption{Variation of the field amplitude versus tortoise
coordinate $x$ with $M=1$, n=1, $\omega=1$, $l=1$,
$\Lambda=10^{-3}$ and $y_{1}=10^{3/2}$ (a very large 5th
dimension).}\label{fig:solution-x}
\end{figure}
\end{center}

\begin{center}
\begin{figure}
\centering
\includegraphics[width=3.5in]{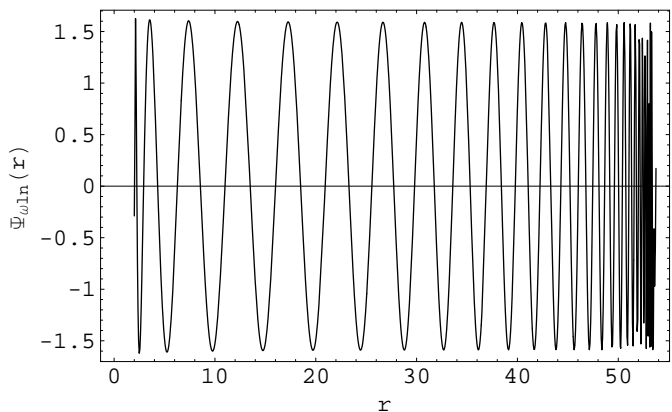}
\caption{Variation of the field amplitude versus $r$
 with $M=1$, n=1, $\omega=1$, $l=1$, $\Lambda=10^{-3}$ and
$y_{1}=10^{3/2}$ (a very large 5th dimension). The waves pile up
near the outer horizon.}\label{fig:solution-r}
\end{figure}
\end{center}

Because such approximation (\ref{tangent-appr}) does not allow
very large $|x|$, we shorten the distance of $x$ to [-10,180], and
hence boundary condition (\ref{Boundary-conditions}) can be
rewritten as
\begin{equation} \label{100-boundary-condition}
    \Psi_{\omega l n}(-10)=\cos(20M\omega),\   \Psi_{\omega l
    n}(180)=\cos(360M\omega).
\end{equation}
By using Mathematica software, combining the
Schr$\ddot{o}$dinger-like equation (\ref{radius-equation}) and
boundary conditions (\ref{100-boundary-condition}), we can solve
Eq. (\ref{radius-equation}) numerically as a boundary value
problem. The amplitude versus the tortoise coordinate is
illustrated in Fig. \ref{fig:solution-x}. It can be seen that the
solution $\Psi_{\omega l n}(x)$ is similar to a harmonic wave
without considering the decay factor $1/r$ in expression
(\ref{wave function}). Taking into account the real surrounding,
we use the tortoise transformation (\ref{tor-grav-sf}) and plot
the amplitude versus $r$ in Fig. \ref{fig:solution-r}. Since the
penetrating power becomes weaker, the wave is sparser than
classical case \cite{Brevik} near the black horizon. From the view
of penetration of a square barrier, the difference can be
demonstrated more clearly in the next section.

\section{The Reflection and Transmission}
We assume that the particle flux with energy $E$ bursts towards a
square well along the positive direction of $x$ axis, where the
potential is
\begin{equation}\label{square-well}
    \hat{V}(x)=\left\{
\begin{array}{c}
V_{0}, \ \ \ \ x_{1}<x<x_{2},\\
\ \ \ 0, \ \ \ x<x_{1}\  \text{or}\  x>x_{2}.\\
\end{array}
\right.
\end{equation}
From the view of quantum mechanics, considering the wave behavior
of the particles, this process is similar to scattering on the
surface of propagation medium with thickness of $|x_{2}-x_{1}|$.
Parts of them are transmitted and parts of them are reflected
back. According to statistical interpretation of wave function,
whether the energy $E > V_{0}$ or not, there is definite
probabilities to transmit and reflect by the potential. The
reflection and transmission coefficients denote the magnitude of
those probabilities.
\begin{figure}[tbph]
  \includegraphics[width=3.5in]{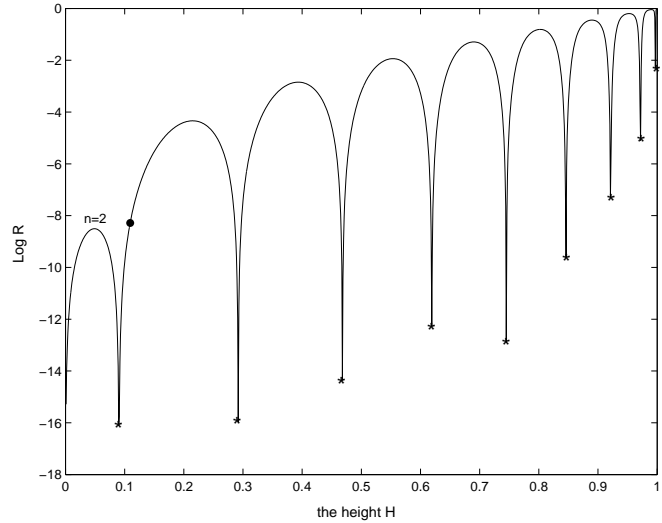}\\
  \caption{The reflected $\log R$ versus the height $H$ (or $\hat{V}_{2}$). The same width $d=14$ with $n=2$ is adopted. Here, $M=1$, $l=1$, $\Lambda=10^{-3}$, and $y_{1}=10^{3/2}$. The resonant points are marked by the star symbol (\text{$*$}). The n=2 case is represented by a black point.}\label{fig:R_H_n2}
\end{figure}

\begin{figure}[btph]
  \includegraphics[width=3.5in]{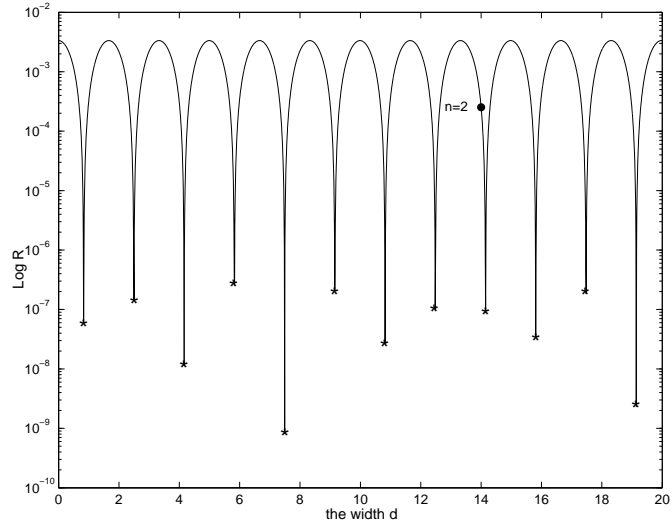}\\
  \caption{The reflected $\log R$ versus the width $d$. The same width $H=0.1093$ with $n=2$ is adopted too. Here, $M=1$, $l=1$, $\Lambda=10^{-3}$, and $y_{1}=10^{3/2}$. The resonant points are marked by the star symbol (\text{$*$}). The n=2 case is represented by a black point.}\label{fig:R_d_n2}
\end{figure}

\begin{figure}
\centering
\includegraphics[width=3.5in]{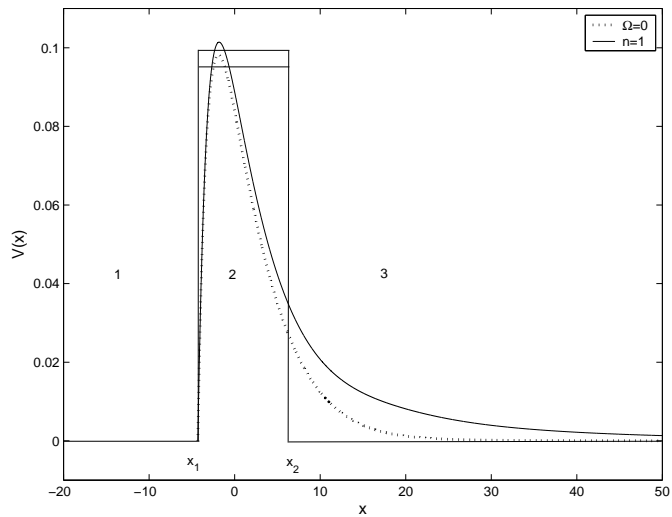}
\caption{Replacement of the real 5D quantum potential barrier
$V_{1}$ (solid) by square barrier with $M=1$, $l=1$,
$\Lambda=10^{-3}$, and $y_{1}=10^{3/2}$ (a very large 5th
dimension). The replacement of the corresponding 4D solution
($\Omega = 0$) is also plotted by dashed line for
comparison.}\label{fig:replacement}
\end{figure}

As mentioned above, it is necessary to replace the continuously
varying potential barrier with a discontinuous barrier of constant
height in analytical work. Therefore, the usual reflection and
transmission coefficients can be obtained. With the method of
\cite{Stratton} and \cite{Brevik}, We suppose a scalar wave
propagates from $-\infty$ to $+\infty$, which is illustrated in
Fig.\ref{fig:replacement}. The same denotation is cited here,
namely associating $''1''$ with the incoming wave in the region
$-\infty<x<x_{1}$, $''2''$ with the potential plateau
$x_{1}<x<x_{2}$, and $''3''$ with the outgoing wave in the region
$x_{2}<x<+\infty$. Hence, potential $V(x)$ in
Eq.(\ref{radius-equation}) reduces to
\begin{equation}\label{sp}
    V(x)=\left\{
\begin{array}{c}
\hat{V}_{1},\ -\infty<x<x_{1},\\
\hat{V}_{2},\ \ \ x_{1}<x<x_{2},\\
\ \ \hat{V}_{3},\ \ \ x_{2}<x<+\infty.\\
\end{array}
\right.
\end{equation}

Following those square barriers, the solutions to the
Eq.(\ref{radius-equation}) are
\begin{equation}\label{solu-p}
    \Psi_{\omega l n}=\left\{
\begin{array}{c}
a_{1}e^{ik_{1} x}+b_{1} e^{-ik_{1} x},\ -\infty<x<x_{1},\\
a_{2} e^{i k_{2} x}+b_{2} e^{-ik_{2} x},\ \ \ x_{1}<x<x_{2},\\
a_{3} e^{ik_{3}x},\ \ \ \ \ \ \ \ \ \ \ \ \ \ \ \ x_{2}<x<+\infty,\\
\end{array}
\right.
\end{equation}
where $k_{i}=\sqrt{4M^2(\omega^2-\hat{V}_{i})}$ ($i=1,2,3$) are
the wave numbers, $a_{i}$ and $b_{i}$ are the undetermined
coefficients to the solutions . Then we define the reflection
coefficients for the plane interfaces dividing two media
\begin{equation}\label{Rij}
    R_{ij} = \left(\frac{1-Z_{ij}}{1+Z_{ij}}\right)^2,
\end{equation}
where $Z_{ij} = \frac{k_{j}}{k_{i}}$ are the real impedance ratios
between medium $i$ and $j$. So the width of the barrier is
$d=x_{2}-x_{1}$ and the height of the square barrier is $H =
\hat{V}_{2}$. So in this model the reflection coefficients $R$ and
transmission coefficients $T$ are given as \footnote{See Ref
\cite{Brevik} for detail.}
\begin{eqnarray}
  R&=&\left|\frac{b_{1}}{a_{1}}\right|^2=\frac{R_{12}+R_{23}+2\sqrt{R_{12}R_{23}}\cos(2k_{2}d)}{1+R_{12}R_{23}+2\sqrt{R_{12}R_{23}}\cos(2k_{2}d)} ,\label{R}\\
 T&=&\left|\frac{a_{3}}{a_{1}}\right|^2=\frac{1}{(1+Z_{12})^2(1+Z_{23})^2}\frac{16}{1+R_{12}R_{23}+2\sqrt{R_{12}R_{23}}\cos(2k_{2}d)}.\label{T}
\end{eqnarray}

It is well known that in quantum mechanics \cite{Zeng}, the
resonant effect frequently occurs in the square barrier. After the
particle flux enters into the square barrier, the particle flux is
reflected and transmitted again both by the walls $x_{2}$ and
$x_{1}$ (Fig. \ref{fig:replacement}). If wavelength satisfies the
resonant condition, the resonant tunnelling would be arisen here.
As a matter of fact, the waves after scattering many times have
the same phasic. Their coherent states are stacked together.
Hence, the amplitude of transmission waves increase evidently.
Naturally, the reflection is weak more. Also, one can read this
feature directly from Eq. (\ref{R}) and Eq. (\ref{T}), which
contain cosine functions. So there is no surprise that Fig.
\ref{fig:R_H_n2} and Fig. \ref{fig:R_d_n2}, which describe $\log
R$ vs. $H$ and $\log R$ vs. $d$ respectively, take oscillating
like forms. When $\log R$ gets the extremum points (stars), the
resonant tunnelling is arisen in the square barrier. At the same
time, the case of $n=2$ (two black points) is presented in the
both figures, too.

\begin{table}[b]
\tabcolsep 0pt \caption{The reflection and transmission
coefficients} \vspace*{-12pt}
\begin{center}
\def\temptablewidth{0.5\textwidth}
{\rule{\temptablewidth}{1pt}}
\begin{tabular*}{\temptablewidth}{@{\extracolsep{\fill}}ccccccc}
mode & \ $x_{1}$\ &\ $x_{2}$\ &\ d\ &\ \ \ \ \ $\hat{V_{2}}$\ \ \
\ \ &$R$ &\ \ \ \ \ $T$\ \ \ \ \ \\\hline
          4D ($\Omega=0$) &\ -4 &\ 6 &\ 10 &\ 0.0950 &\ $2.4\times10^{-3}$ &\
          0.9975\\
          5D ($n=1$) &\ -4 &\ 6 &\ 10 &\ 0.1000 &\ $2.7\times10^{-3}$ &\
          0.9973\\
         5D ($n=2$) &\ -4 &\ 10 &\ 14 &\ 0.1093 &\ $2.5\times10^{-4}$ &\ 0.9997 \\
         5D ($n=3$) &\ -4 &\ 20&\ 24 &\ 0.1267 &\ $1.9\times10^{-3}$ &\
         0.9981
       \end{tabular*}\label{table}
       {\rule{\temptablewidth}{1pt}}
       \end{center}
       \end{table}

In order to get reflection and transmission coefficients in this
5D black string, we have to find the approximative square
potentials to replace the continuous potentials $V_{n}$. So the
replacements is plotted in Fig. \ref{fig:replacement}. For
clearness of the figure, here we only consider the 4D ($\Omega=0$)
and the first 5D quantum potential ($n=1$).
In the case of $n=1$, we read $\hat{V_{2}}=0.1000$ as a reasonable
value for the potential plateau in Fig. \ref{fig:replacement}, and
choose the incoming wave number to be $k_{1}=2$ $(\hat{V_{1}}=0)$
and $k_{3}=2$ $\hat{(V_{3}}=0)$. For the case of 4D ($\Omega=0$),
we employ the same width and read $\hat{V_{2}}=0.0950$. Meanwhile,
using the same method, we can get the width and altitude of square
barrier ($n=2$) and ($n=3$), respectively. In the end, every
parameter in those modes is given in Table (\ref{table}).
Substituting those into Eqs. (\ref{R}) and (\ref{T}), the
reflection and transmission coefficients ($R$, $T$) are obtained
and listed in Table (\ref{table}) too. Because of the resonant
effect, it is difficult to judge directly whether R (or T, notice
R+T=1) would be smaller or larger with the change of n only by
considering  the width $d$ or the height $H$.  Careful contrast of
the Fig. \ref{fig:R_H_n2} and Fig \ref{fig:R_d_n2} shows up some
key differences that although they all oscillate integrality, the
average order of $R$ varies obviously with the height $H$ rather
than the width $d$.

\section{conclusion }
In this paper we have solved the real scalar field $\Psi_{\omega l
n}$ and obtained reflection coefficients and transmission
coefficients in the 5D black string space. We summarize what has
been achieved, and make some further comments.

1. The 5D black string solution presented by Mashhoon, Liu and
Wesson in Ref. \cite{Mashhoon} is exact in higher dimensional
gravitional theory. In the work of paper \cite{Liu00}, the
Klein-Gordon equation has been successfully separated into three
parts, $\Phi \sim R_{\omega}(r,t)L(y)Y_{lm}(\theta,\phi)$,
corresponding to the variables ($r,t$), the fifth coordinate $y$,
and the usual spherical harmonic coordinates ($\theta, \varphi$).
We notice that the extra dimension $y$ or $\xi$ affects the usual
4D by the parameter $\Omega$, which is contained in the potential
(\ref{potential-of-r}). The wave function $L(y)$ gets its extremum
on two branes. Then by using the steady standing wave condition,
the quantum potential $V_{n}$ are obtained after quantizing the
parameter $\Omega$. So the effective 5D potential
(\ref{quan-potential}), which is different from the usual 4D case,
determines the evolution of massless particles around the black
hole (note that when $\Omega=0$, the potential
(\ref{potential-of-r}) reduce to the usual 4D Schwarzschild-de
Sitter black hole potential \cite{Brevik}). Naturally, we can get
a new radial wave function $\Psi_{\omega l n}$ being homologic
with $\Psi_{\omega l}$ \cite{Brevik} in 4D.

2. One purpose of this paper is to find the difference between
$\Psi_{\omega l n}$ and $\Psi_{\omega l}$. We study scalar
particles scattering around our black string solution
(\ref{eq:5dmetric}) in the bulk and get the scalar field solution
under the tangent approximation. Eq. (\ref{radius-equation})
describes the evolution of one dimensional transmission of a wave
through the potential barrier. According to Eq.
(\ref{radius-equation}), effective potential
(\ref{quan-potential}) and boundary condition
(\ref{Boundary-conditions}), a full boundary value problem is
presented. Because of the complex potential (\ref{quan-potential})
and the approximative replacement (\ref{tangent-appr}), we only
give the numerical solution instead of a analytical one.

3. From the view of mathematics, an oscillatory cosine function
$\cos(2k_{2}d)$ or $\cos( 2 \sqrt{ 4M^2 (\omega^2 - H )} d)$ is
contained both in $R$ (\ref{R}) and $T$ (\ref{T}). It is obviously
that the function $R(H,d)$ and $T(H,d)$ are not monotonic
function. We can see it clearly from Fig. \ref{fig:R_H_n2} and
Fig. \ref{fig:R_d_n2}. Considering the different replacement of
square barrier obtained through the different quantum potentials,
we only show the case n=2 (black points)in the figures. It is
obvious that it is not easy to distinguish which one is smaller or
bigger just by $H$ or $d$. Since that our replacement is a
approximate method and its width and height are adopted very
roughly, it is not necessary to give this resonant condition.

4. It is known that the hierarchy problem, which is why the
characteristic scale of gravity ($M_{p}\sim 10^{9}$ GeV) is about
16 orders of magnitude larger than the Electro-Weak scale
($W_{EW}\sim 1$ TeV), is solved successfully by assuming the
existence of extra dimensions both in ADD model \cite{Arkani}
\cite{Antoniadis} \cite{Shiu} and R-S model \cite{Randall1}
\cite{Randall2}. It is considered that the extra dimensions would
be really visible at the neighborhood of black holes and in the
early epoch of universe . If it does so,, regarding the former,
the extra dimensionality of space would emerge out. In the strong
gravitational field
--- black hole or black string (this model), the best way to probe
this is the Hawking radiation. It contains a lot of information
such as the mass of black hole or black string, the topological
structure of our brane-world, the magnitudes of effective
cosmological constant $\Lambda$ and so on. The appearance of wave
function $\Psi_{\omega l n}$ is purely a quantum effect of the
fifth dimensions. In this paper we employ a approximative method
to give the real scalar field solution in the bulk roughly.

5. In this paper we only consider the quantum number n depended
property of scalar field. Unlike the work \cite{Kanti3}, we do not
take into account the dependence on the effective cosmological
constant $\Lambda$. We take $\Lambda = constant$ instead. The
reason is that in order to solve the Sch\"{o}dinger-like Eq.
(\ref{radius-equation}), which has two coordinates variables $r$
and $x$, we need to find a quasi function (\ref{1}). Nevertheless,
we have to choose the value of effective cosmological constant
firstly in the key function (\ref{tor-grav-sf}). If the value of
$\Lambda$ varies, the approximate function has to be changed too.
Furthermore, it is difficult to find out the approximate relation
between the radial coordinate $r$ and the tortoise coordinate $x$.
So in this work, the previous value $\Lambda = 10^{-3}$
\cite{Brevik} \cite{Liu00} and the tangent approximation
(\ref{tangent-appr}) is employed. Anyway, it is interesting to
research this case and the further work is needed.

\acknowledgments This work was supported by NSF (10573003) and
NBRP (2003CB716300) of P. R. China. Xu was supported in part by
DUT 893321.

\end{document}